\documentclass[12pt]{article}

\usepackage{amsmath}
\usepackage{amssymb}
\usepackage{latexsym}
\usepackage{graphicx}
\usepackage{epsfig}

\def\beq{\begin{equation}}
\def\eeq{\end{equation}}
\def\bea{\begin{eqnarray}}
\def\eea{\end{eqnarray}}

\begin{document}
\begin{titlepage}

\vspace*{1cm}

\begin{center}
{\bf {\Large {A new approach to information loss (no) problem  \\[2mm] for Black Holes}}}

\bigskip \bigskip \medskip

{\bf Nikolaos D. Pappas}

\bigskip

{\it Division of Theoretical Physics, Department of Physics,\\
University of Ioannina, Ioannina GR-45110, Greece}

\bigskip \bigskip \bigskip \medskip
{ \bf{Abstract}}
\end{center}

    The discovery that black holes emit thermal type radiation changed
    radically our perception of their behavior. Until then, their interior
    was considered as causally disconnected from the rest of the universe,
    so any kind of information, that went down the black hole, was believed
    to remain eternally trapped in it. The emission of the aforementioned
    radiation means that some amount of information eventually returns to
    the universe outside the black hole. The question then rises whether it
    is the whole of this information that goes back to the universe during
    the black hole evaporation or not. Numerous theories supporting either
    information preservation or extinction have been developed ever since.
    A new idea is proposed, based on a deep re-examination of what
    information is and what are its properties. We postulate that not all
    kinds of information are of equal importance to nature and, as a result,
    some of them should be preserved under any conditions, while the rest
    are allowed to be destroyed, so both preservation and destruction of
    information is what actually happens during the black hole formation/evaporation
    process.

\bigskip\bigskip\bigskip\bigskip\bigskip\bigskip\bigskip\bigskip\bigskip\bigskip\bigskip\bigskip
\begin{center}
{ \it{E-mail: npappas@cc.uoi.gr}}
\end{center}

\end{titlepage}

\setcounter{page}{1} \noindent

\section{Introduction}

More than 30 years ago Hawking proved, using a semiclassical
approximation, that black holes radiate because of the inevitable
creation of pairs of particles due to quantum energy fluctuations
at the vicinity of their horizon. Furthermore, he was able to show
that this radiation is exactly thermal, that is no subtle or
secret correlations exist between the emitted particles
\cite{Hawking1, HH}. As we know nowadays, black holes actually
behave like black bodies of a specific temperature T given by the
equation
\begin{eqnarray}
T = \frac{c^{3}\hbar}{8{\pi}k_{B}GM}
\end{eqnarray}
Or, in terms of solar mass $M_{\odot}$,
\begin{eqnarray}
T=6.2\times10^{-8}(\frac{M_{\odot}}{M})K
\end{eqnarray}

Hawking radiation drew much attention since then as, for the very
first time, physicists dealt with a procedure which results from
the combination of a purely quantum mechanical process, such as
particle creation from vacuum, with the dynamical properties of
space-time, that are governed by the laws of General Relativity.
Space-time is no longer considered as the passive background where
quantum phenomena take place, but, on the contrary, its curvature
and the existence of an event horizon are indispensable in order for a
black hole to emit particles. Nevertheless, even quantum phenomena
must abide by the energy conservation principle. Hence, when
Hawking radiation escapes to infinity, we may safely conclude that
it will carry energy away from the black hole, which must
therefore shrink in mass. As the mass shrinks the surface gravity
increases and with it the temperature. This is a self-catalyst
process in which the entire mass evaporates away in a finite time.

For astrophysical black holes of mass M this procedure is
particularly slow and their lifetime, although finite, is very
long, calculated to be of the order
\begin{eqnarray}
\tau_{BH}\sim(\frac{M}{m_{pl}})t_{pl}\sim
(\frac{M}{M_{\odot}})^{3}10^{71}sec
\end{eqnarray}
where $m_{pl}\thickapprox10^{-5}gr$ is the Planck
mass and $t_{pl}\thickapprox10^{-43}sec$ is the Planck time. All
these mean that the estimated lifetime of a solar mass black hole
is $10^{53}$ times larger than the current age of our universe,
whose Hubble time is $t_{H}=H_{0}^{-1}\varpropto10^{18}sec$. The
duration of the process may seem extremely long for the
evaporation of black holes to have any practical impact on us, but
one should notice that the lifetime of mini black holes is so much
shorter that primordial black holes could reach the end of their
lives today. Even more, since we are talking about the very
principles of Physics, no Gedankenexperiment can be regarded as
extreme enough so as not to be worth considering.

The evaporation, that comes as a result of the aforementioned
radiation, posed an unexpected question. Initially, the interior
of black holes used to be considered as causally disconnected from
the rest of the universe, so any kind of information, that gets
inside a black hole, was believed to remain eternally trapped in
it. We know that this is not the case, as black holes radiate,
and, therefore, some amount of information eventually returns back
to our universe. The major question, accompanying this
observation, is whether, during the formation/evaporation process
of a black hole, information is preserved or gets partially
destroyed.

Every scientist, that concerns him/herself with this issue,
before anything else, always bumps into the question whether information is actually preserved or destroyed. Whatever the
answer, the effort to support it gives rise to new questions and
challenges. If information is conserved, then one should propose
or invent some kind of a mechanism, which ensures this. Bearing in
mind that we lose track of some amount of information in every
ordinary process, why should black holes preserve it in the first
place? If, on the other hand, one accepts the possibility that
information can be destroyed, questions about when and why this
happens should be addressed. Being it the case, what makes a black
hole to abstain from destroying the whole of it and lose even its
classical hair? And for the whole matter to get more complicated,
unitarity, a key demand of Quantum Theory, appears to be violated
in the context of the formation and evaporation of black holes
since particles in pure state, that get absorbed by them, end up
in mixed state as parts of Hawking radiation.

All these questions outline the celebrated information loss
problem, which has been tantalizing physicists for more that three
decades. This problem actually consists of two quite different
issues. The first one could be described as ``loss of history''
meaning that two black holes of the same mass, charge and angular
momentum radiate exactly the same way, even though they probably
have absorbed different objects during their lifetime. Therefore,
we lose knowledge about the specific properties of whatever goes
down a black hole apart from the three parameters mentioned above.
The second one has to do with the apparent non-unitary evolution
of particles that black holes seem to evoke, as stated earlier.
Here we will only deal with the ``loss of history'' problem and
leave the salvation of unitarity to be the subject of a future
work.

There are numerous papers, where theories about possible
preservation mechanisms are presented by several scientists, since
most of the physicists find the idea of information destruction
and the subsequent breakdown of predictability to be unpalatable.
The most significant ones have to do with the invention of some
mechanism through which an enormous amount of information can
either be encoded in Hawking radiation \cite{Hawking2, BekeMuk,
SHW, Hod, Muk,VSK} or is forced to remain trapped in the inaccessible
interior of a black hole remnant \cite{ACN}. Both of them, though,
are still far from being considered as complete solution to the
problem, since they have serious drawbacks. The existence of
complicated but subtle correlations in the spectrum of Hawking
radiation possible as it may seem, meaning, of course, that the latter
is nearly and not exactly thermal, consists a deviation from our present 
knowledge that also
has to be explained. Furthermore, for the equilibrium between
ingoing and outgoing information to hold, Hawking radiation should
carry a really huge amount of all kinds of information and it is
quite hard for one to see how this could be realized by these
alleged correlations that, in any case, are assumed to be very
feeble (for a convincing presentation of the arguments undermining
the validity of such a solution see \cite{Mathur} and for a proposed 
way out see \cite{VS}). As far as
black hole remnants are concerned, their existence is even more
problematic as their abundance and total mass are calculated to be
so large, that their gravitational impact on the known universe
should have already been detected. Not to mention that only vague
assumptions can be made about what kind of mechanism can stop the
evaporation and save black holes from extinction by creating an
extremely stable and long-lived remnant, that is left behind at
the end \cite{Beke1, Suss}.

Quite a few other theories have been proposed over the years to
address the problem, where information comes out massively once
the black hole reaches the Planck size when the semiclassical
approximation is no longer trustworthy \cite{Page}, escapes into a
baby-universe \cite{FGSA}, is conserved in space-times of $1+1$
dimensions \cite{ATV, PST}, remains trapped inside the infinitely
large interior of cornucopions (a variation of remnants)
\cite{BOS, Banks, Giddings1, BDDO}, is stored in a topologically disconnected
from our Universe region, created inside the black hole due to a topological
change process that the horizon undergoes spontaneously, the latter being a
fuzzy sphere in the first place \cite{Silva} etc. Despite any virtues they
may have, these theories suffer from very serious defects that
make them least viable (see for example \cite{Preskil} for an
extensive and thorough presentation and analysis of various
theories both mainstream and exotic ones, and also
\cite{Giddings2}).

Significantly fewer papers appear in the literature to support the
possibility of information destruction \cite{Preskil, Hawking3},
since the issue is not whether it can get lost for all practical
purposes, but if it can be destroyed in a way that is irreversible
in principle. Most of the physicists seem reluctant to defend such
a prospect, as one would have to answer why and how this could
happen and what are the limitations of this procedure.
Nevertheless, our present knowledge implies that this is probably
the case even if this means that we should alter or expand some of
our ideas concerning how nature works.

\section{Classes of information}

Up till now, it must be evident that the information loss paradox
cannot be encountered using only well established and understood
rules and theories. Our current lack of knowledge about the laws
of Quantum Gravity means that every approach is incomplete to some
extend as it is doomed to overlook the fact that the particles
spend a part of their lives interacting with the singularity since
this interaction can only be understood and calculated in the
context of exactly these yet ill understood laws. An innovative
idea is proposed in this paper, which can be described in a
nutshell as ``information classification'', that could be proven to
be a useful tool in the effort to fully understand the nature of
black holes.

We postulate that all kinds of information, in general, about a
physical system fall into two categories. Let's call them $\Pi1$
and $\Pi2$ ($\Pi$ is the first letter of the greek word that
stands for information). $\Pi1$ class contains the most
fundamental information that defines a particle, such as the mass,
the electric and magnetic charge and the angular momentum of it.
The much larger $\Pi2$ category includes information about how
particles mingle with each other and the properties that rise from
their combinations. A book, for example, contains a vast number of
$\Pi1$-info about the aforementioned conserved quantities, of
the elementary particles it consists of, and an even larger number
of $\Pi2$-info about how this particles unite to form different
nucleons, atoms and molecules including also the way all these
combine to form letters and words that mean something.

There is a variety of well known conservation laws that impose the
preservation of $\Pi1$-info by any physical system. Black holes
satisfy this requirement by emitting all kinds of particles that
carry away exactly this type of information. It should be pointed
out here that whether information about leptonic and baryonic
numbers or any fermionic degrees of freedom, in general, should be
considered as $\Pi1$ type is an open question. Although we know
that black holes have no well-defined leptonic or baryonic number
\cite{Hartle, Teitelboim, Beke2}, we cannot tell yet if they
behave in a way that results to the conservation of fermionic
quantum numbers or some combination of them. As for the $\Pi2$
category, there are no such laws to prevent this information from
extinction and, consequently, different processes destroy
different amounts of it. Actually, we conjecture that the various
kinds of $\Pi2$-info resist their destruction to different
degrees. In general, more violent procedures destroy more of them,
but not in a proportional way. For instance, going back to the
book example, tearing it apart leads to some $\Pi2$-info loss,
like the meaning of the sentences and the words written in it. By
burning it, we destroy much more $\Pi2$-info, since now words
and letters disappear and no paper or ink survives, but the atoms,
it was made of, are still there. Because all usual phenomena are
confined to a low energy scale, not even all $\Pi2$-info gets
destroyed during them. This means that in every day life we get to
observe loss only of some part of the total $\Pi2$ category
practically in every process, which leads to an increase of the
total entropy, as dictated by the second law of thermodynamics and
no paradoxes occur. However, the case of black holes is somewhat
different in that, being the most extreme objects in the universe,
the matter, they absorb, undergoes impacts of arbitrarily large
violence, so they destroy all $\Pi2$-info ruthlessly and, as a
result, their entropy adds up to enormous values. Meanwhile $\Pi1$-info
remains intact even in this case.

In a few words, we argue that the concept of information loss as
the physical basis of the second law of thermodynamics
\cite{DunSem} should be completed with the limitation that this
loss can only concern $\Pi2$-info, no matter how easy or
difficult it is to be destroyed. Furthermore, black holes are
postulated to be the most efficient $\Pi2$-info destroyers in
the Universe and no such info can survive after having crossed the
horizon. Rephrasing a well known aphorism by R. Price \cite{Price},
"any $\Pi2$ type piece of information that could be destroyed by a
black hole, will be destroyed".

\section{Information paradox resolution}

Taking under consideration the idea of information classification,
we gain a more incisive perception of the behavior of black holes,
which has significant advantages and almost no flaws. To be more
specific there are six arguments in favor of the new idea.

First,information, as far as the subset $\Pi1$ of fundamental
importance is concerned, is never lost and, therefore, no paradox
rises at this level.

Second, because information does get destroyed by the black hole,
even if it can only be of $\Pi2$ category, the absorption and
incorporation of matter by the black hole is a thermodynamically
favored procedure, as expected by the fact that this is what
always happens in reality.

Third, since black holes interact with the rest of the universe
via their horizon, the rate they absorb matter and, consequently,
destroy the information contained in it, must be proportional to
their surface A. This observation provides us with a possible
explanation of why black hole entropy is directly analogous to
their surface, as explicitly shown by the famous Bekenstein -
Hawking formula
\begin{eqnarray}
 S=\frac{A}{4} \hspace{2em} with \hspace{1em} c=G=\hbar=1
\end{eqnarray}
 and not to their volume, as could one
instinctively assume accounting black holes to be some sort of
ordinary thermodynamical systems.

Fourth, everything we know so far about the behavior of black
holes and the laws governing it, like the Generalized Second Law,
remain intact and valid since no revision of them is necessary in
the context of our theory.

Fifth, it provides us with some mechanism capable of explaining
why black hole entropy generally takes enormous values, as we can
ascribe it to the complete and irreversible destruction of all
$\Pi2$-info.

Sixth, the fact that less information contributes to the creation
of the entropy of black holes, since they destroy only
$\Pi2$-info, even though $\Pi2$ class constitutes the greater part
of the total information load of incoming matter, can help us deal
with the problematic current estimation that the entropy of
ordinary black holes is almost equally large as the entropy of the
Universe. More specifically, the latter is approximately equal to
the number of relativistic particles whose number, within a Hubble
radius, is calculated to be $S_{U}=10^{88}$. On the other hand,
black hole entropy measured in Planck units is equal to the 1/4 of
their surface and converting this quantity into astrophysical
units we find the Bekenstein - Hawking entropy to have the huge
value
\begin{eqnarray}
S_{BH}\sim10^{90}(\frac{M}{10^{6}M_{\odot}})^{2}
\end{eqnarray}

This equation means that a single million solar mass black hole
has more entropy than the whole known universe. At the same time
General Relativity certainly allows for the existence of such
supermassive objects and astronomical observations imply that they
must exist in the centre of almost every galaxy, so another
paradox rises. A possible way out is that the idea of information
classification also means that the way black hole entropy is
calculated should be reconsidered on the basis of $\Pi1$ and
$\Pi2$ information classes in order to express the fact that only
$\Pi2$-info gets destroyed. The refined calculation should result
to an entropy value several orders of magnitude smaller than the
one mentioned above, resolving the new paradox.

Finally, information classification theory is quite appealing in
that, while predicting some kind of information conservation,
avoids any assumptions about remnants or exotic encoding in the
spectrum of Hawking radiation or other even more radical but least
possible alternatives. Its simplicity should be seen as an extra
advantage of it, if one is to trust Occam and his famous
razor \footnote{Occam's razor: philosophical argument stating that 
among all theories, capable of explaining a phenomenon, the one with 
the simplest conjectures is most probably the correct one.}.

Of course, it is still an open question what actually happens to
unitarity, demanded by Quantum Theory, during the
formation/evaporation process of black holes for which the
information classification has nothing new to propose. As
mentioned before, our idea only deals with the ``loss of history''
part of the paradox and, in our opinion, seems to work very
well indeed.

\section{Conclusions}

Information loss paradox is a very interesting and still open
issue that indicates the limits of the $20^{th}$ century physics
and waits for its solution to be found, hopefully, in the
$21^{st}$ century. Several efforts have been made so far to
resolve it, but all of them have such serious drawbacks, that
hardly any can be considered as a good start for a definitive
answer. In this paper, a differentiation in the way we perceive
the notion of information is proposed, based on the assumption
that not all kinds of information are equally important to nature.
We argue that some of them are of fundamental and others of
secondary importance, that are characterized as $\Pi1$ and $\Pi2$
respectively, on the basis that the first ones are protected by a
series of conservation laws against destruction, while the latter
ones are allowed to be destroyed with different degrees of ease.
Postulating that black holes radiate away all $\Pi1$ information
through Hawking radiation and, at the same time, they destroy all
of the $\Pi2$ one, we manage to avoid any paradoxes, while their
behavior remains compatible with the second law of thermodynamics.
In this way, one can claim that the information loss paradox, as
far as the ``loss of history'' is concerned, is resolved, at least
in principle.

\end{document}